\documentclass{article}
\usepackage{spconf,amsmath,graphicx,hyperref}

\usepackage{amsfonts}
\usepackage{makecell}
\usepackage{multirow}
\usepackage{graphicx}
\usepackage{subfig}

\title{EEND-SAA: Enrollment-Less Main Speaker Voice\\Activity Detection using Self-Attention Attractors}
\name{
    Wen-Yung Wu \quad
    Pei-Chin Hsieh \quad
    Tai-Shih Chi\thanks{This research is supported by the National Science and Technology Council, Taiwan under Grant NSTC 113-2221-E-A49-146.}
}
\address{
    Institute of Electrical and Computer Engineering, National Yang Ming Chiao Tung University, Taiwan
}
\begin{document}
%
\maketitle
\begin{abstract}
Voice activity detection (VAD) is essential in speech-based systems, but traditional methods detect only speech presence without identifying speakers.
Target-speaker VAD (TS-VAD) extends this by detecting the speech of a known speaker using a short enrollment utterance, but this assumption fails in open-domain scenarios such as meetings or customer service calls, where the main speaker is unknown. We propose EEND-SAA, an enrollment-less, streaming-compatible framework for main-speaker VAD, which identifies the primary speaker without prior knowledge. Unlike TS-VAD, our method determines the main speaker as the one who talks more steadily and clearly, based on speech continuity and volume. We build our model on EEND using two self-attention attractors in a Transformer and apply causal masking for real-time use. Experiments on multi-speaker LibriSpeech mixtures show that EEND-SAA reduces main-speaker DER from 6.63\% to 3.61\% and improves F1 from 0.9667 to 0.9818 over the SA-EEND baseline, achieving state-of-the-art performance under conditions involving speaker overlap and noise.
\end{abstract}
\section{Introduction}
\label{sec:intro}

Voice activity detection (VAD) \cite{sohn1999statistical} is a core module in smart voice systems, detecting speech and non-speech regions and serving as the front-end for automatic speech recognition, speaker diarization, and voice agents. Traditional methods using energy thresholds or zero-crossing rate (ZCR) \cite{schafer1975digital, moattar2009simple} are efficient but fail in low SNR or overlapping speech, while statistical approaches like GMMs and HMMs \cite{park2004voice, ronao2014human} improve robustness only under moderate noise.

With deep learning, VAD evolved to end-to-end neural networks. CNNs capture local time-frequency patterns \cite{svirsky2023sg, hershey2017cnn}, dilated causal convolutions handle longer contexts \cite{chang2018temporal}, RNNs and LSTMs model temporal dependencies \cite{lee2020dual}, and Transformers exploit self-attention for global context under noisy, overlapping conditions. Target-speaker VAD (TS-VAD) \cite{ding2022personal, cheng2023target, makishima2021enrollment} further detects a known speaker using an enrollment utterance \cite{desplanques2020ecapa}, but it relies on prior knowledge and fails in open-domain scenarios.

In real-world cases such as call centers, talks, or meetings, enrollment is often impractical. We focus on main-speaker VAD (MS-VAD), which directly infers the dominant speaker using cues like volume and continuity, making it suitable for multi-speaker scenarios without enrollment.

Speaker diarization also segments speech and clusters unknown speakers without enrollment. Traditional pipelines run VAD, extract embeddings, cluster via Agglomerative Hierarchical Clustering (AHC) \cite{day1984efficient}, and refine with HMMs \cite{fox2011sticky}, but suffer from overlap, varying speaker counts, and high latency. End-to-End Neural Diarization (EEND) \cite{fujita2019end} predicts speaker activity directly using BiLSTM with permutation invariant training (PIT) \cite{yu2017permutation}. SA-EEND \cite{fujita2019endsa} replaces BiLSTM with Transformer \cite{vaswani2017attention} for better inter-speaker modeling, and EEND-EDA \cite{nagamatsu2020end} introduces attractors for variable speaker counts. However, EEND variants only predict “who spoke when” and cannot label the main speaker, and require full sequences, making them unsuitable for streaming.

To address these limitations, we propose EEND-SAA, a streaming-compatible, enrollment-less framework for main-speaker VAD. We extend the EEND backbone by introducing Dual Self-Attention Attractors, which split the attention heads into two parts: one focuses on the main speaker, and the other on background speakers. This lets the system answer two questions: whether there is speech, and whether it belongs to the main speaker. We also apply causal masking and key-value caching during inference for real-time, low-latency processing, making the system more suitable for interactive voice systems in noisy environments.

Our main contributions are:
\begin{itemize}
\item We formalize enrollment-less main-speaker VAD, offering a practical solution for multi-speaker, noisy conditions without pre-recorded reference speech.
\item We redesign the attractor module using a self-attention structure, improving main-speaker tracking under dynamic scenarios.
\item We incorporate an adaptive causal mechanism for real-time operation, demonstrating potential for smart voice agents and interactive systems.
\end{itemize}

\begin{figure}[t]
\centering
\includegraphics[width=0.8\columnwidth]{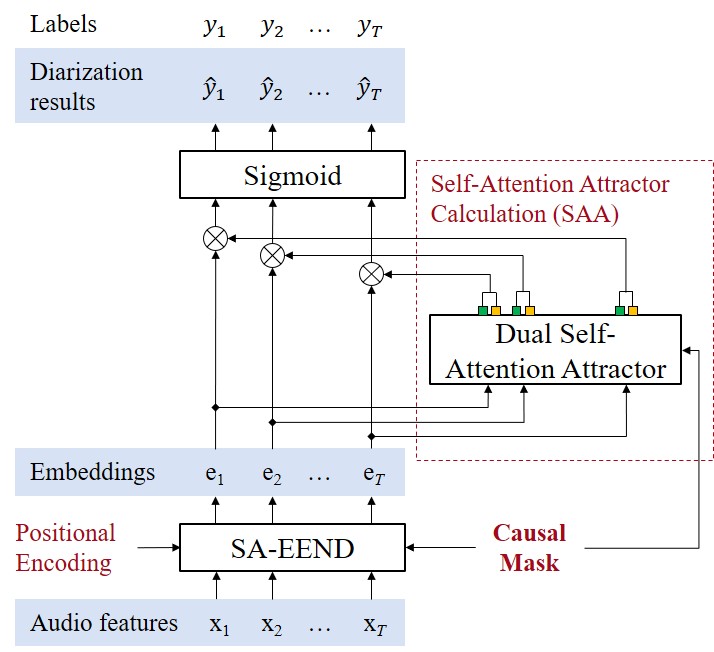} 
\caption{Architecture of EEND-SAA with causal awareness}
\label{fig:EEND-SAA}
\end{figure}

\section{Methods}

\subsection{Model Architecture}

The proposed EEND-SAA architecture is inspired by SA-EEND and EEND-EDA, with the introduction of self-attention in the attractor generation stage. It consists of three modules: an EEND encoder, the Self-Attention Attractor (SAA) module, and the frame-level speaker activity output module, as shown in Figure~\ref{fig:EEND-SAA}.

The EEND encoder converts the input log-Mel spectrogram into a $T \times D$ sequence of speaker-aware embeddings. A Transformer models temporal context and overlapping speech using multi-head self-attention to capture long-range dependencies. Unlike SA-EEND, we add positional encoding to provide temporal awareness, helping detect continuity in the main speaker’s speech, which tends to be steadier than others. SA-EEND omits this to mimic clustering behavior where order is irrelevant, but for main-speaker VAD, temporal order improves precision. The positional encoding is defined as:

\begin{equation}
\begin{aligned}
  \mathrm{PE}(pos,2i)   &= \sin\!\Bigl(\frac{pos}{10000^{2i/d}}\Bigr) \\
  \mathrm{PE}(pos,2i+1) &= \cos\!\Bigl(\frac{pos}{10000^{2i/d}}\Bigr)
\end{aligned}
\end{equation}
where $\text{pos}$ is the position index, $i$ is the embedding dimension index, and $d$ is the model dimension. This design links temporal position with speaker behavior, improving main-speaker detection.


The SAA module is central to our system. Inspired by EEND-EDA, it replaces the LSTM-based encoder-decoder with a self-attention mechanism for better long-range modeling. After multi-head attention, time-averaging produces speaker-specific attractors, which are compared with input embeddings to detect main-speaker activity. To further separate main and background speakers, SAA adopts a dual attention structure, splitting attention heads: one focuses on the main speaker, the other on background speakers, generating two distinct attractors and improving role distinction.

After computing the attractors, the embedding sequence $E \in \mathbb{R}^{T \times D}$ and attractors $A \in \mathbb{R}^{2 \times D}$ are used to compute the speaker activity probability for each frame:

\begin{equation}
\begin{aligned}
\hat{Y} = \sigma(E A^\top),
\end{aligned}
\end{equation}
where $\hat{Y}$ is the frame-by-speaker probability matrix and $\sigma$ is the element-wise sigmoid. The model outputs probabilities for the main and background speakers per frame.






\begin{table*}[!ht]
\caption{speaking and silence configuration}
\label{table:voice_silence}
\centering
\begin{tabular}{c|c|c|c|c|c}
\hline
configuration index & configuration target & \makecell{start delay \\ (sec)} & \makecell{voice duration \\ \(t_{\text{voice}}\)(sec)} & \makecell{silence duration \\ \(t_{\text{silence}}\)(sec)} & proportion of voice \\ \hline
M0 & \makecell{main speaker / \\ background speakers} & \multirow{5}{*}{0-3} & 7-10 & N/A & N/A \\ \cline{1-2} \cline{4-6}
B1 & \multirow{4}{*}{background speakers} &  & 1-3 & 0-6 & 40.0\% \\ \cline{1-1} \cline{4-6}
B2 &  &  & 1-4 & 0-5 & 50.0\% \\ \cline{1-1} \cline{4-6}
B3 &  &  & 1-4 & 0-3 & 62.5\% \\ \cline{1-1} \cline{4-6}
B4 &  &  & 1-4 & 0-1 & 83.3\% \\ \hline
\end{tabular}
\end{table*}

\subsection{Causal Streaming Architecture}
Real-time streaming in EEND-SAA is achieved by applying causal masking to both the Transformer and SAA modules, ensuring predictions depend only on current and past inputs. Unlike the original full-sequence averaging, the causal version generates attractors per time step and directly compares them with the current embedding.  Key-value caching further reduces redundant computation for low-latency inference.

To handle cases where background speakers precede the main speaker, we adopt a causal-aware labeling strategy for training. Early background speech is temporarily treated as main-speaker activity, and background labels are omitted during this period. Once the main speaker begins, the labels are correctly assigned back to the true main speaker, allowing the model to learn how to gradually shift attractor focus based on continuity and volume. Combined with the causal design, the system achieves efficient and accurate real-time performance in multi-speaker environments.

\section{Experiments}

\subsection{Dataset and Data Generation}
We used the LibriSpeech corpus to synthesize training data with overlapping speakers, noise, and turn-taking to simulate real-world conditions. Each 15-second sample was mixed from 2–4 randomly selected speakers, with one randomly assigned as the main speaker. Background speech followed predefined alternating voice and silence intervals $t_{\text{voice}}$ and $t_{\text{silence}}$ in Table~\ref{table:voice_silence}, simulating fragmented patterns to increase overlap. Background volume was varied using scaling ranges [0.1, 0.4], [0.2, 0.8], and 1.0 to add natural variability.

To enhance robustness, MUSAN noise \cite{snyder2015musan} (office, street, ambient) and random Room Impulse Responses (RIRs) were added per speaker to simulate reverberant environments.  Raw audio was downsampled to 8 kHz, transformed via STFT, converted to Mel filter banks, log-compressed, and mean-normalized to produce the log-Mel spectrogram as model input.

In total, 100k training samples were generated from 921 speakers in \texttt{train-clean-360}, and 1k samples each were built for validation and test sets from \texttt{dev-clean} and \texttt{test-clean}, covering all mixing configurations.

\subsection{Model Configuration and Training}


The input feature is the log-Mel spectrogram, computed with a 25 ms window, 10 ms hop, and 23 Mel filters. Features are normalized per utterance and subsampled by retaining every third frame, which is concatenated with its three preceding and succeeding frames to preserve context while reducing sequence length.

The EEND encoder uses four Transformer layers with four attention heads, 256-dimensional embeddings, a 2048-dimensional feed-forward network, and 0.1 dropout rate for training. The SAA module has one self-attention layer with four heads, while Dual-SAA doubles heads to eight, splitting them for main and background speakers.

The model was trained for 100 epochs using a batch size of 64 with Adam optimizer and a Noam learning rate schedule:

\begin{equation}
\begin{aligned}
\text{lr} = d_{\text{model}}^{-0.5} \cdot \min \big(& \text{step\_num}^{-0.5}, \\
& \text{step\_num} \cdot \text{warmup\_steps}^{-1.5} \big)
\end{aligned}
\end{equation}

where $d_{\text{model}}=256$, and we set 100k warm-up steps for stable training.

A weighted BCE loss is used during training, with factor $\alpha$ controlling the contribution of the main speaker loss $L_{\text{main}}$ and background speaker loss $L_{\text{others}}$.
The overall objective is:

\begin{equation}
\begin{aligned}
L_{\text{BCE}} &= L_{\text{main}} + \alpha L_{\text{others}}
\end{aligned}
\end{equation}

\begin{table}[t]
\caption{Results of the original SA-EEND on the standard segmentation task with different numbers of speakers}
\label{table:sa-eend_spk_comparison}
\centering
\begin{tabular}{cccc}
\hline
speakers \# & DER (\%) & $\text{DER}_\text{main}$ (\%) & $\text{F1}_\text{main}$ \\ \hline
2 & 18.27 & 8.44 & 0.9561 \\ 
3 & 24.31 & 10.59 & 0.9451 \\ 
4 & 30.31 & 14.91 & 0.9244 \\ \hline
\end{tabular}
\end{table}

\begin{table}[t]
  \caption{Model comparison on the test data set}
  \label{table:model_comparison}
  \centering
  \resizebox{\columnwidth}{!}{
    \begin{tabular}{lccc}
      \hline
      model
        & DER (\%)
        & $\mathrm{DER}_\mathrm{main}$ (\%)
        & $\mathrm{F1}_\mathrm{main}$ \\
      \hline
      \makecell{SA-EEND \cite{fujita2019endsa}\\(w/ 2–4 speakers labels)}     & 25.84 & 11.44 & 0.9412 \\
      \makecell{SA-EEND \cite{fujita2019endsa}\\(w/ main speaker labels)}      & N/A   &  6.63 & 0.9667 \\
      EEND-SAA (single)                         & N/A   &  4.21 & 0.9788 \\
      EEND-SAA (dual)                         &  7.46 &  \textbf{3.61} & \textbf{0.9818} \\
      \hline
    \end{tabular}
  }
\end{table}

\subsection{Evaluation Metrics}

We evaluate the system using Diarization Error Rate (DER), Main Speaker DER (DER\textsubscript{main}), and Macro F1. DER is defined as:

\begin{equation}
\text{DER} = \frac{N_{\text{Miss}} + N_{\text{FA}} + N_{\text{Confusion}}}{N_{\text{Total}}}
\end{equation}

where $N_{\text{Miss}}$, $N_{\text{FA}}$, and $N_{\text{Confusion}}$ represent missed speech, false alarms, and speaker confusion, respectively. DER\textsubscript{main} evaluates only the main speaker by excluding speaker confusion, while Macro F1 equally weighs each test sample to handle imbalanced data and unknown speakers.



\section{Experimental Results}

\subsection{Comparison of Architectures}

We compare four variants: (1) original SA-EEND, (2) SA-EEND trained with main-speaker labels, (3) EEND-SAA with a single attractor, and (4) EEND-SAA with dual attractors. All models used the same setup: background config B2 (50\% speaking ratio), volume scaling 0.2--0.8, and 2–4 speakers per sample.

Table~\ref{table:sa-eend_spk_comparison} shows that SA-EEND suffered higher DER and DER\textsubscript{main} as speaker count increased, due to its fixed-speaker assumption and lack of main-speaker awareness.
As shown in Table~\ref{table:model_comparison}, training SA-EEND with main-speaker labels reduced DER\textsubscript{main} and improved F1\textsubscript{main}.  Adding the self-attention attractor enabled dynamic main-speaker localization, and the dual-attractor design further improved accuracy by contrasting background speech and suppressing irrelevant segments.

\begin{table}[t]
  \caption{EEND-SAA results of different speech ratios in the test set with the background speaker volume scaling of 1/0.2-0.8}
  \label{table:different_voice_silence_1}
  \centering
    \begin{tabular}{cccc}
      \hline
      \makecell{speaking\\config}
        & DER (\%)
        & $\mathrm{DER}_\mathrm{main}$ (\%)
        & $\mathrm{F1}_\mathrm{main}$ \\
      \hline
      M0   & 13.52/9.12 & 18.53/8.38 & 0.9063/0.9582 \\ 
      B1   &  8.46/8.83 &  \textbf{5.84}/\textbf{3.40} & \textbf{0.9707}/\textbf{0.9828} \\ 
      B2   &  7.50/7.46 &   5.92/3.61 & 0.9702/0.9818 \\ 
      B3   &  6.89/7.20 &   7.33/3.99 & 0.9634/0.9799 \\ 
      B4   & \textbf{6.49}/\textbf{6.17} &   8.42/4.35 & 0.9579/0.9781 \\ 
      \hline
    \end{tabular}
\end{table}

\begin{table}[t]
\caption{Results of EEND-SAA under different volume scaling in the test set of background speaker configuration M0/B2}
\label{table:different_volume_result_1}
\centering
    \begin{tabular}{cccc}
    \hline
    \makecell{volume\\scaling}
        & DER (\%)
        & $\mathrm{DER}_\mathrm{main}$ (\%)
        & $\mathrm{F1}_\mathrm{main}$ \\
      \hline
    0.1-0.4 & 10.43/8.99 & \textbf{5.39}/\textbf{2.51} & \textbf{0.9729}/\textbf{0.9873} \\
    0.2-0.8 & \textbf{9.12}/\textbf{7.46} & 8.38/3.61 & 0.9582/0.9818 \\
    1 & 13.52/7.50 & 18.53/5.92 & 0.9063/0.9702 \\ \hline
    \end{tabular}
\end{table}


\subsection{Analysis on Speech Activity Ratio}

We analyzed how different speaking ratios affect main-speaker detection while keeping volume fixed. Background volume was set to 1.0, and speaking/silence intervals varied from continuous speech (M0) to highly intermittent (B4). As shown in Table~\ref{table:different_voice_silence_1}, F1\textsubscript{main} dropped to  about 0.90 in M0 due to confusion with continuous background speech, but rose above 0.97 in B1--B4, showing that the model can leverage temporal stability to identify the main speaker.

We further tested robustness with background volume randomly scaled between 0.2 and 0.8. Table~\ref{table:different_voice_silence_1} shows that DER\textsubscript{main} remained stable (3.40\%–4.35\%) even with more active background speech, confirming the model’s reliance on continuity.

Interestingly, when background activity increased, overall DER decreased while DER\textsubscript{main} slightly rose, indicating that active backgrounds aid diarization but create more competition for main-speaker detection. Still, the model maintained strong accuracy by tracking the main speaker’s patterns.

\subsection{Impact of Background Volume}

We further examined how background volume affects detection under fixed speaking patterns. In either M0 or B2 setting, lowering volume to 0.1--0.4 significantly raised F1\textsubscript{main} and reduced DER\textsubscript{main}, as shown in Table~\ref{table:different_volume_result_1}, confirming that volume helps distinguish speakers even with the same speech timing setting. It indicates with softer background speech, the model more clearly detected the main speaker.


Results in Table~\ref{table:different_voice_silence_1} and~\ref{table:different_volume_result_1} show that both volume and continuity guide segmentation. Even with louder background speech, the model remained robust by focusing on the main speaker’s consistent patterns, and vice versa with volume.  This confirms adaptability in real-world conditions.

\begin{table}[!t]
\caption{Testing results of causal model}
\label{table:causal_results}
\centering
\resizebox{\columnwidth}{!}{
    \begin{tabular}{lcccc}
    \hline
    model & \makecell{causal\\data} & DER (\%) & $\text{DER}_\text{main}$ & $\text{F1}_\text{main}$ (\%) \\ \hline
    Dual SAA & No & \textbf{7.46} & 3.61 & 0.9818 \\
    Dual SAA w/ causal & No & 11.34 & 10.39 & 0.9512 \\
    Dual SAA w/ causal & Yes & 8.17 & \textbf{3.28} & \textbf{0.9835} \\ \hline
    \end{tabular}
}
\end{table}

\begin{table}[t]
\caption{Comparison of EEND-SAA results with and without positional encoding}
\label{table:pos_results}
\centering
    \begin{tabular}{cccc}
    \hline
    model & DER (\%) & $\text{DER}_\text{main}$ (\%) & F1 \\ \hline
    EEND-SAA w/ pos & \textbf{7.46} & \textbf{3.61} & \textbf{0.9818} \\ 
    EEND-SAA w/o pos & 10.51 & 7.45 & 0.9626 \\ \hline
    \end{tabular} 
\end{table}

\subsection{Causal Model}

In real-time applications, the model must work without future context. We adapted EEND-SAA into a causal version and designed matching labels for this constraint.
Table~\ref{table:causal_results} shows that applying a causal model on non-causal data caused F1\textsubscript{main} to drop significantly, and DER\textsubscript{main} rose to 10.39\%. This confirms that mismatch between the model and the label design reduces performance.  With causal-aware labels, the model starts predicting main-speaker activity even while background speakers begin speaking first, and then shifts focus once the main speaker begins. This improves response time and boundary accuracy in streaming tasks.

\subsection{Ablation Study: Effect of Positional Encoding}

We evaluated the impact of positional encoding by comparing EEND-SAA with and without it, as shown in Table~\ref{table:pos_results}.  Without position information, the model struggled to track speech flow and often misclassified background speech as main-speaker activity.  Adding positional encoding improved sequence modeling, enabling the attractor to better identify the main speaker, showing that temporal order provides useful cues even without explicit speaker identity.

    
    
    

\section{Conclusion}
We present a main-speaker voice activity detection model that integrates EEND with SAA, enabling accurate main speaker identification in overlapping and noisy speech without enrollment.  By leveraging attractor-based modeling and temporal structures, the model shows strong robustness to background interference and maintains high accuracy when the main speaker is continuous or background speakers are intermittent.  With causal masking and adapted labeling, the system also supports real-time inference, enhancing its practical applicability.

\bibliographystyle{IEEEbib}
\bibliography{strings,refs}

\begin{thebibliography}{10}

\bibitem{sohn1999statistical}
Jongseo Sohn, Nam~Soo Kim, and Wonyong Sung,
\newblock ``A statistical model-based voice activity detection,''
\newblock {\em IEEE signal processing letters}, vol. 6, no. 1, pp. 1--3, 1999.

\bibitem{schafer1975digital}
Ronald~W Schafer and Lawrence~R Rabiner,
\newblock ``Digital representations of speech signals,''
\newblock {\em Proceedings of the IEEE}, vol. 63, no. 4, pp. 662--677, 1975.

\bibitem{moattar2009simple}
Mohammad~Hossein Moattar and Mohammad~Mehdi Homayounpour,
\newblock ``A simple but efficient real-time voice activity detection algorithm,''
\newblock in {\em 2009 17th European signal processing conference}. IEEE, 2009, pp. 2549--2553.

\bibitem{park2004voice}
Kiyoung Park, Changkyu Choi, and Jeongsu Kim,
\newblock ``Voice activity detection using global soft decision with mixture of gaussian model.,''
\newblock in {\em INTERSPEECH}, 2004, pp. 965--968.

\bibitem{ronao2014human}
Charissa~Ann Ronao and Sung-Bae Cho,
\newblock ``Human activity recognition using smartphone sensors with two-stage continuous hidden markov models,''
\newblock in {\em 2014 10th international conference on natural computation (ICNC)}. IEEE, 2014, pp. 681--686.

\bibitem{svirsky2023sg}
Jonathan Svirsky and Ofir Lindenbaum,
\newblock ``Sg-vad: stochastic gates based speech activity detection,''
\newblock in {\em ICASSP 2023-2023 IEEE International Conference on Acoustics, Speech and Signal Processing (ICASSP)}. IEEE, 2023, pp. 1--5.

\bibitem{hershey2017cnn}
Shawn Hershey, Sourish Chaudhuri, Daniel~PW Ellis, Jort~F Gemmeke, Aren Jansen, R~Channing Moore, Manoj Plakal, Devin Platt, Rif~A Saurous, Bryan Seybold, et~al.,
\newblock ``Cnn architectures for large-scale audio classification,''
\newblock in {\em 2017 ieee international conference on acoustics, speech and signal processing (icassp)}. IEEE, 2017, pp. 131--135.

\bibitem{chang2018temporal}
Shuo-Yiin Chang, Bo~Li, Gabor Simko, Tara~N Sainath, Anshuman Tripathi, A{\"a}ron van~den Oord, and Oriol Vinyals,
\newblock ``Temporal modeling using dilated convolution and gating for voice-activity-detection,''
\newblock in {\em 2018 IEEE international conference on acoustics, speech and signal processing (ICASSP)}. IEEE, 2018, pp. 5549--5553.

\bibitem{lee2020dual}
Joohyung Lee, Youngmoon Jung, and Hoirin Kim,
\newblock ``Dual attention in time and frequency domain for voice activity detection,''
\newblock in {\em Proc. Interspeech 2020}, 2020, pp. 3670--3674.

\bibitem{ding2022personal}
Shaojin Ding, Rajeev Rikhye, Qiao Liang, Yanzhang He, Quan Wang, Arun Narayanan, Tom O’Malley, and Ian McGraw,
\newblock ``Personal vad 2.0: Optimizing personal voice activity detection for on-device speech recognition,''
\newblock in {\em Proc. Interspeech 2022}, 2022, pp. 3744--3748.

\bibitem{cheng2023target}
Ming Cheng, Weiqing Wang, Yucong Zhang, Xiaoyi Qin, and Ming Li,
\newblock ``Target-speaker voice activity detection via sequence-to-sequence prediction,''
\newblock in {\em ICASSP 2023-2023 IEEE International Conference on Acoustics, Speech and Signal Processing (ICASSP)}. IEEE, 2023, pp. 1--5.

\bibitem{makishima2021enrollment}
Naoki Makishima, Mana Ihori, Tomohiro Tanaka, Akihiko Takashima, Shota Orihashi, and Ryo Masumura,
\newblock ``Enrollment-less training for personalized voice activity detection,''
\newblock in {\em Proc. Interspeech 2021}, 2021, pp. 346--350.

\bibitem{desplanques2020ecapa}
Brecht Desplanques, Jenthe Thienpondt, and Kris Demuynck,
\newblock ``Ecapa-tdnn: Emphasized channel attention, propagation and aggregation in tdnn based speaker verification,''
\newblock in {\em Proc. Interspeech 2020}, 2020, pp. 3830--3834.

\bibitem{day1984efficient}
William~HE Day and Herbert Edelsbrunner,
\newblock ``Efficient algorithms for agglomerative hierarchical clustering methods,''
\newblock {\em Journal of classification}, vol. 1, no. 1, pp. 7--24, 1984.

\bibitem{fox2011sticky}
Emily~B Fox, Erik~B Sudderth, Michael~I Jordan, and Alan~S Willsky,
\newblock ``A sticky hdp-hmm with application to speaker diarization,''
\newblock {\em The Annals of Applied Statistics}, pp. 1020--1056, 2011.

\bibitem{fujita2019end}
Yusuke Fujita, Naoyuki Kanda, Shota Horiguchi, Kenji Nagamatsu, and Shinji Watanabe,
\newblock ``End-to-end neural speaker diarization with permutation-free objectives,''
\newblock in {\em Proc. Interspeech 2019}, 2019, pp. 4300--4304.

\bibitem{yu2017permutation}
Dong Yu, Morten Kolb{\ae}k, Zheng-Hua Tan, and Jesper Jensen,
\newblock ``Permutation invariant training of deep models for speaker-independent multi-talker speech separation,''
\newblock in {\em 2017 IEEE International Conference on Acoustics, Speech and Signal Processing (ICASSP)}. IEEE, 2017, pp. 241--245.

\bibitem{fujita2019endsa}
Yusuke Fujita, Naoyuki Kanda, Shota Horiguchi, Yawen Xue, Kenji Nagamatsu, and Shinji Watanabe,
\newblock ``End-to-end neural speaker diarization with self-attention,''
\newblock in {\em 2019 IEEE Automatic Speech Recognition and Understanding Workshop (ASRU)}. IEEE, 2019, pp. 296--303.

\bibitem{vaswani2017attention}
A~Vaswani,
\newblock ``Attention is all you need,''
\newblock {\em Advances in Neural Information Processing Systems}, 2017.

\bibitem{nagamatsu2020end}
Kenji Nagamatsu, Yawen Xue, Yusuke Fujita, Shota Horiguchi, and Shinji Watanabe,
\newblock ``End-to-end speaker diarization for an unknown number of speakers with encoder-decoder based attractors,''
\newblock {\em Interspeech 2020}, 2020.

\bibitem{snyder2015musan}
David Snyder, Guoguo Chen, and Daniel Povey,
\newblock ``Musan: A music, speech, and noise corpus,''
\newblock {\em arXiv preprint arXiv:1510.08484}, 2015.

\end{thebibliography}

\end{document}